\documentclass[a4paper,onecolumn,11pt,accepted=2021-09-21]{quantumarticle}
\pdfoutput=1

\usepackage[utf8]{inputenc}
\usepackage[T1]{fontenc}
\usepackage{amsthm, amsmath, amssymb}
\usepackage[numbers,sort&compress]{natbib}
\usepackage{booktabs}
\usepackage{graphicx}
\usepackage{braket}
\usepackage{epstopdf}
\usepackage{algorithmicx}
\usepackage[noend]{algpseudocode}
\usepackage{algorithm}
\usepackage{ytableau}
\usepackage[caption=false]{subfig}
\usepackage[colorlinks]{hyperref}
\usepackage[all]{hypcap}

\algrenewcommand\algorithmicrequire{\textbf{Precondition:}}
\algrenewcommand\algorithmicensure{\textbf{Postcondition:}}
\usepackage{scalerel}    
\usepackage{listings}    
\usepackage{hyperref}
\usepackage{bm}

\usepackage[normalem]{ulem}

\usepackage{xcolor}

\definecolor{codegreen}{rgb}{0,0.6,0}
\definecolor{codegray}{rgb}{0.5,0.5,0.5}
\definecolor{codepurple}{rgb}{0.58,0,0.82}
\definecolor{backcolour}{rgb}{0.95,0.95,0.92}
\lstdefinestyle{mystyle}{
  backgroundcolor=\color{backcolour},   commentstyle=\color{codegreen},
  keywordstyle=\color{magenta},
  numberstyle=\tiny\color{codegray},
  stringstyle=\color{codepurple},
  basicstyle=\fontsize{7.8}{7.8}\selectfont\ttfamily,
  breakatwhitespace=false,         
  breaklines=true,                 
  captionpos=b,                    
  keepspaces=true,                 
  showspaces=false,                
  showstringspaces=false,
  showtabs=false,
  xleftmargin=0.02\textwidth,
  rulecolor=\color[RGB]{200,200,200},
  frame=bt,
  framextopmargin=5pt,
  framexbottommargin=5pt,
  framexleftmargin=10pt,
  tabsize=2
}
\begin{document}
\title{The Fermionic Quantum Emulator}

\date{10/15/2021}
\author{Nicholas C. Rubin}
\email[Corresponding author: ]{nickrubin@google.com}
\affiliation{Google Quantum AI, Mountain View, CA, 94043}
\author{Klaas Gunst}
\affiliation{Quantum Simulation Technologies, Inc., Cambridge, MA 02139}
\author{Alec White}
\affiliation{Quantum Simulation Technologies, Inc., Cambridge, MA 02139}
\author{Leon Freitag}
\affiliation{Quantum Simulation Technologies, Inc., Cambridge, MA 02139}
\author{Kyle Throssell}
\affiliation{Quantum Simulation Technologies, Inc., Cambridge, MA 02139}
\author{Garnet Kin-Lic Chan}
\email[Corresponding author: ]{garnetc@caltech.edu}
\affiliation{Division of Chemistry and Chemical Engineering, California Institute of Technology, Pasadena
CA 91125}
\author{Ryan Babbush}
\email[Corresponding author: ]{babbush@google.com}
\affiliation{Google Quantum AI, Mountain View, CA, 94043}
\author{Toru Shiozaki}
\email[Corresponding author: ]{shiozaki@qsimulate.com}
\affiliation{Quantum Simulation Technologies, Inc., Cambridge, MA 02139}

\begin{abstract}
The fermionic quantum emulator (FQE) is a collection of protocols for emulating quantum dynamics of fermions efficiently taking advantage of common symmetries present in chemical, materials, and condensed-matter systems.  The library is fully integrated with the OpenFermion software package and serves as the simulation backend. The FQE reduces memory footprint by exploiting number and spin symmetry along with custom evolution routines for sparse and dense Hamiltonians, allowing us to study significantly larger quantum circuits at modest computational cost when compared against qubit state vector simulators. This release paper outlines the technical details of the simulation methods and key advantages.
\end{abstract}
\maketitle
\section{Introduction}
High accuracy simulation of fermionic systems is an important and challenging problem and is a major motivation behind current attempts to develop  quantum computers~\cite{Feynman1982,Abrams1997,Ortiz2001,Aspuru-Guzik2005}.
There has been significant experimental progress towards realizing the simulation of fermionic systems on current quantum devices~\cite{rubinScience2020, kandala2017hardware}. As these experiments scale in size there is a growing need to understand the possibilities for quantum advantage, with one approach being to characterize the classical emulation complexity of the corresponding quantum circuits. In addition, the efficient emulation of near-term fermionic simulation experiments is crucial for experiment design, algorithm design, and testing.
In this work, we describe an implementation of protocols to  efficiently emulate quantum circuits describing time evolution under fermionic generators. We name the library that implements these protocols the Fermionic Quantum Emulator (FQE).\footnote{We make a distinction between simulating quantum circuits directly and emulating the quantum circuits when the quantum circuit corresponds to dynamics of fermions.}

There have been many developments in quantum circuit simulation and emulation.  Broadly these advancements can be classified into algorithmic improvements~\cite{Markov2008, bravyi2019simulation, PhysRevA.103.022428, PhysRevA.99.052307,gray2021hyper, boixo2017simulation, gray2021hyper, huang2020classical} and computational implementation improvements~\cite{smelyanskiy2016qhipster, luo2020yao, suzuki2020qulacs, qforte}. However, despite this progress, there remains potential to improve the emulation of circuits relevant to fermionic simulation. For example, in general circuit emulation  it is necessary to include additional circuit elements to handle the fermion encoding, but these can be eliminated and the fermionic sign accounted for implicitly in a specialized fermionic emulator. 
In addition, many fermionic systems have symmetries that can be used to reduce the resource requirements of the classical emulation. For example, molecular and material problems are often described by Hamiltonians
that commute with a variety of global symmetry operators, such as the total particle number, total spin, time-reversal, and point-group and crystallographic symmetries. 
Though there are many open source software packages to
carry out quantum chemistry calculations of fermionic systems using these symmetries~\cite{sun2018pyscf, smith2020psi4, shiozaki2018bagel}, these are not designed to support computations using the quantum circuit model. 
Similarly, existing quantum circuit simulators and the corresponding computational techniques used within them 
are not naturally suited to efficiently working within
symmetry reduced Hilbert spaces.

The FQE is a simulator of fermions represented in second quantization. It corresponds to a statevector simulator in that the wavefunction is explicitly stored. In the first release of the emulator, number and spin ($S_z$) symmetry are used to minimize the wavefunction memory footprint by using a generalization of the Knowles--Handy~\cite{Knowes1984CPL} determinant based configuration interaction wavefunction organization scheme. Time evolution of the state under a fermionic Hamiltonian is implemented in two ways: (1) via a custom routine similar to the cosine-sine construction for nilpotent operators applicable to sparse fermion generators, and (2) via a series expansion of the propagator for dense Hamiltonians taking advantage of efficient intermediate data structures. The library completely integrates with the fermion quantum simulation library OpenFermion~\cite{mcclean2020openfermion} and has built in functions to convert wavefunction objects in FQE to the format used in OpenFermion as well as for the general purpose circuit simulator Cirq~\cite{cirq_developers_2021_4586899}.  For example, the time evolution of a wavefunction can be generated using OpenFermion's \texttt{FermionOperator} objects or Hamiltonians defined through FQE utility functions.  The FQE is an open source library distributed under the Apache 2.0 license~\cite{fqe_2020}.  The library is implemented in Python to facilitate code extension and reuse. However, some hot spots have been addressed with high-performance kernels written in the C language. These C extensions can be enabled or disabled by the user. This allows the library to function as either a high-performance library for computationally demanding applications or as a pure-Python library implemented with understandable and easily modifiable expressions.

This work describes the key technical aspects of the library, demonstrates how it can be used in quantum algorithm development, and makes single-thread and multi-threaded timing comparisons for key circuit primitives against Cirq~\cite{cirq_developers_2021_4586899} and the highly performant general purpose quantum circuit simulator Qsim~\cite{quantum_ai_team_and_collaborators_2020_4023103}. We close with a perspective on the development of the library and future directions.

\section{Wavefunctions and Hilbert space organization}
When simulating spin-$\frac{1}{2}$ fermions in second quantization, the many-particle fermionic Hilbert space (sometimes called the Fock space) generated by a basis of $L$ single-fermion states (termed orbitals) $\{ |\phi_1\rangle, |\phi_2\rangle, \ldots |\phi_L\rangle \}$, is spanned by
basis states (termed determinants) labelled by $L$-digit binary strings $\{n_1 n_2 \ldots n_L\}$ where $n \in \{ 0, 1\}$, and $n_i$ is the occupancy of orbital $i$. It is evident that the fermionic Hilbert space is isomorphic to the Hilbert space of $L$ qubits, with the qubit computational basis corresponding to the fermion determinant basis. This allows one to use the traditional state vector representation of qubits to encode fermionic states. 

For many molecular and materials systems, the electronic Hamiltonian is an operator that commutes with
various global symmetry  operators. These symmetries thus provide useful quantum numbers to label sectors of Hilbert space. 
Because the Hamiltonian does not mix sectors, many  simulations can be performed either in a single sector, or in a collection of independent sectors. A simple example is simulating a fixed number of electrons $n$. In this case, the Hilbert space contains only ${L}\choose{n}$ states. Because the determinants are eigenstates of the particle number operator, the spanning basis can be chosen to be determinants with total occupancy $\sum_{i} n_{i} =n$. If we further assume the orbitals are eigenstates of $S_{z}$ (i.e. spin-orbitals), then the determinants are also eigenstates of $S_{z}$. Then a Hilbert space sector labelled by $n$ and $S_z$ is spanned by determinants only with the given $n$ and $S_z$. 

The FQE uses such a decomposition into symmetry sectors to store the wavefunction compactly. A user specifies symmetry sectors of fixed particle number and given $S_{z}$. The total wavefunction is then stored in the direct sum of these sectors.  Considering all possible particle sectors and $S_{z}$ sectors corresponds to working in the full many-particle fermionic Hilbert space, and thus storing the full set of $2^{L}$ qubit amplitudes. 

An efficient way to represent wavefunctions for a fixed particle number and $S_z$ sector was first introduced by Siegbahn \cite{Siegbahn1984CPL} and refined by Knowles and Handy~\cite{Knowes1984CPL} in the context of the exact diagonalization of chemistry Hamiltonians. The binary string encoding a determinant is separated into a  string for the spin-up ($\alpha$)
spin-orbitals and the spin-down ($\beta$) spin-orbitals. Commonly these two binary strings (integers) are 
referred to as the $\alpha$- and $\beta$-string respectively. A further simplification arises by assuming that the $\alpha$ and $\beta$ spin-orbitals share a common set of spatial functions, known as spatial orbitals. Given $M$ spatial orbitals in total, the total number of spin-orbitals is thus $L=2M$. 

The following code example initializes a state with four electrons in four spatial orbitals over a superposition of all possible $S_{z}$ values. 
\begin{lstlisting}[language=Python]
wfn = fqe.Wavefunction([[4, 4, 4], [4, 2, 4], [4, 0, 4], [4, -2, 4], [4, -4, 4]])
\end{lstlisting}
Each element of the input list labels a Hilbert space sector as the triple ($n$-electrons,  $S_{z}$, $M$-spatial-orbitals). The wavefunction in each sector is stored as a matrix of size ${M}\choose{n_{\alpha}}$ $\times$ ${M}\choose{n_{\beta}}$ where $n_\alpha$ and $n_\beta$ are the total number of $\alpha$ and $\beta$ electrons.
Each row of the matrix  corresponds to a specific $\alpha$-string,  and each column  to a $\beta$-string.
The mapping between the integer values of the $\alpha$- and $\beta$-strings to the row and column indices is somewhat arbitrary; we use the lexical ordering proposed by 
Knowles and Handy \cite{Knowles1989CPC}. By storing the wavefunction coefficients as matrices we can leverage vectorized multiplications when performing updates or contractions on the wavefunction coefficients. 

\begin{figure}
    \centering
    \includegraphics[width=7.0cm]{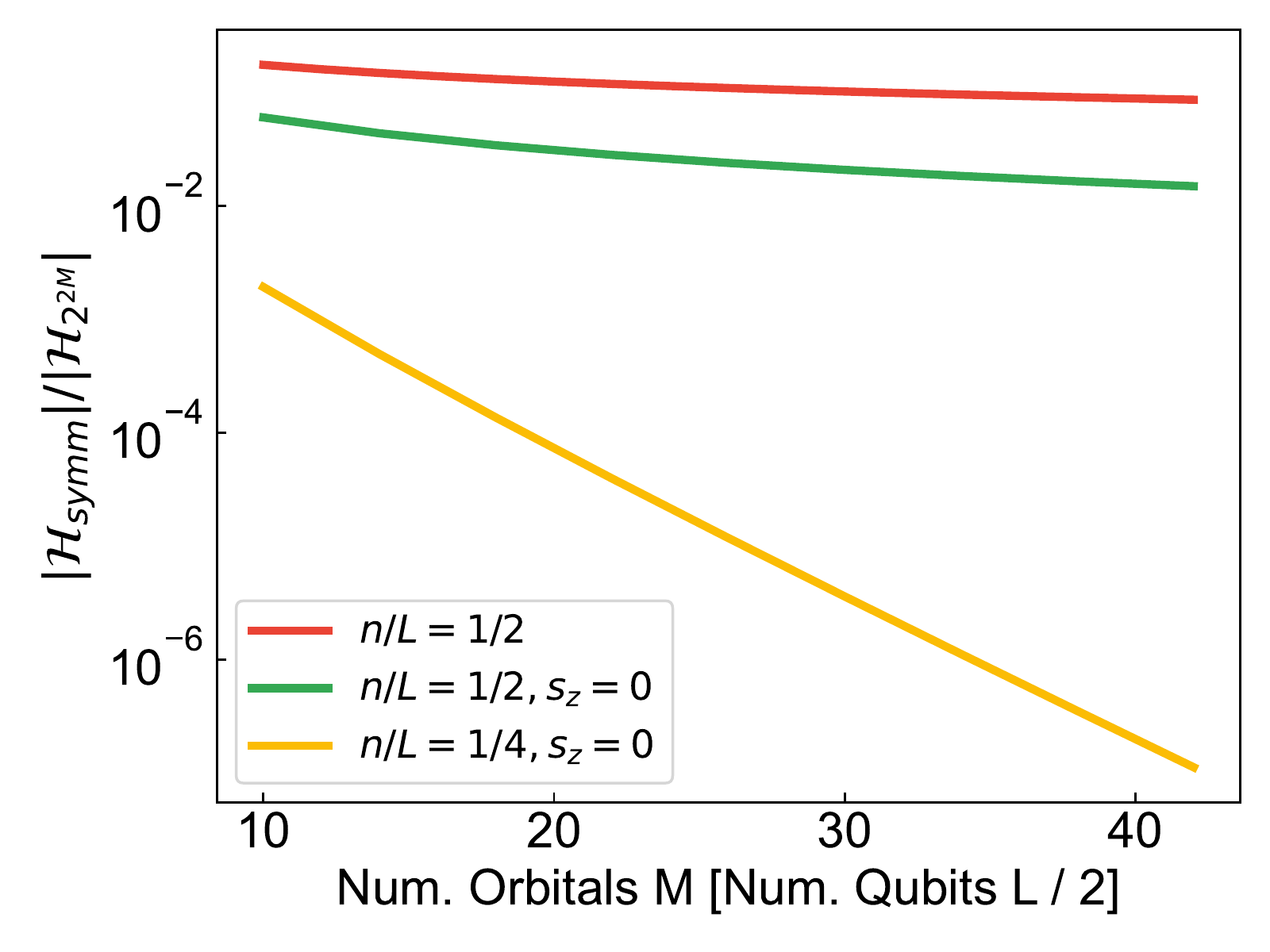}
    \includegraphics[width=7.0cm]{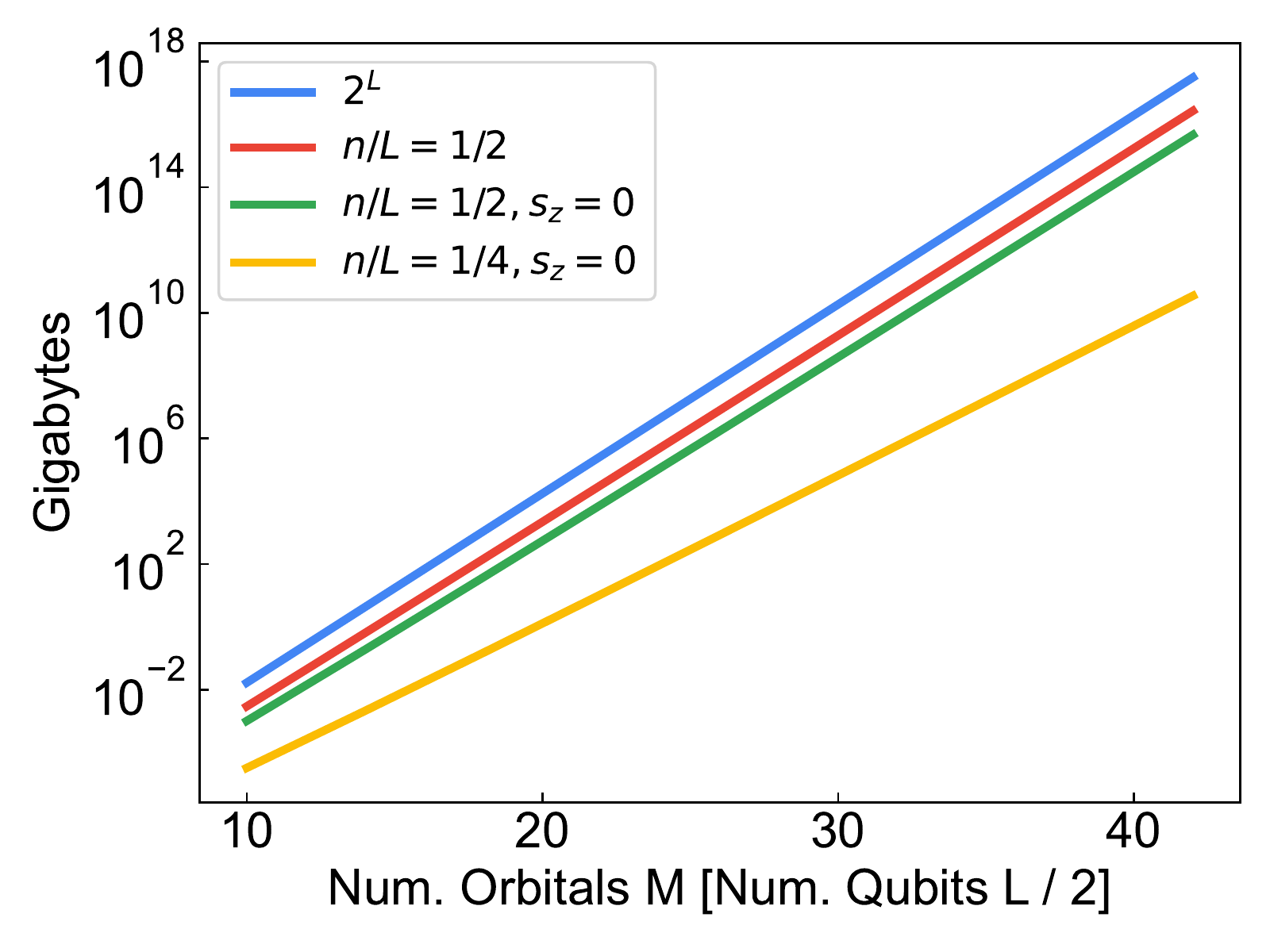}
        \caption{\textit{left:} Relative sizes of Hilbert space sectors at half and quarter filling compared against full qubit Hilbert space size. \textit{right:} Gigabytes required to represent spaces of $S_z=0$, half and quarter filling, compared against the full qubit Hilbert space and half filling with no $S_{z}$ symmetry restriction.}
    \label{fig:wf_size}
\end{figure}

The memory savings from simulating in specific symmetry sectors is substantial when compared to traditional state-vector simulators which use the full $2^L$ qubit space.
In Figure~\ref{fig:wf_size} we show the relative size of half filling ($n=L/2$) and quarter filling ($n=L/4$) subspaces with $S_{z}=0$ along with the wavefunction memory footprint in gigabytes.

To make the FQE interoperable with other simulation tools we provide a number of utilities for transforming the FQE \texttt{Wavefunction} representation.  We also provide human readable printing, saving wavefunctions as binary data in \texttt{numpy}'s \texttt{.npy} format, various wavefunction initialization routines for random, Hartree-Fock, or user defined states, and conversion functions to OpenFermion and Cirq wavefunction representations.  The printing functionality is demonstrated in the code snippet below.
\\
\\
\\
\begin{lstlisting}[language=Python]
import fqe

# two sectors N=2 Sz=[0,-2] on 4 orbitals
wf = fqe.Wavefunction([[2, 0, 4], [2, -2, 4]]) 
wf.set_wf(strategy='random')
wf.print_wfn()
\end{lstlisting}
\begin{lstlisting}
Sector N = 2 : S_z = -2
a'0000'b'0011' (0.10858096183738326-0.34968321927438234j)
a'0000'b'0101' (-0.09372404351124913-0.19775900492450152j)
... elided output
Sector N = 2 : S_z = 0
a'0001'b'0001' (0.09251611820297777+0.1450093827514917j)
a'0001'b'0010' (0.15516111875500693+0.17752453798314619j)
a'0001'b'0100' (0.12499026886480313+0.028546904174259435j)
... elided output
\end{lstlisting}
The conversion to OpenFermion is handled by converting each 
coefficient in the 
\texttt{Wavefunction} object 
with its associated $\alpha$, $\beta$-string
to an ordered string of creation operators acting on a vacuum state, with the string represented by an OpenFermion  \texttt{FermionOperator} object. This allows the user to leverage the normal ordering routines in OpenFermion. By using the \texttt{FermionOperator} intermediate we can also directly map the string of normal ordered operators acting on a vacuum to a state vector for integration with Cirq simulations.
\begin{lstlisting}[language=Python]
of_ops = fqe.fqe_to_fermion_op(wf)

cirq_wf = fqe.to_cirq(wf)
new_fqe_wf = fqe.from_cirq(cirq_wf, thresh=1.0E-12)  # same as original wf
\end{lstlisting}

\section{Unitary evolution \label{section:unitary-evolution}}
\subsection{Evolution of Sparse Hamiltonians}
The FQE can evolve a state by any fermionic excitation Hamiltonian.  A fermionic excitation Hamiltonian is of the form
\begin{align}
\hat{H}_{\text{excite}} =& \epsilon\left(\hat{g} + \hat{g}^{\dagger}\right) \\
\hat{g} =& g\prod^{N_\mathrm{op}}_{p=1}\hat{a}_{i_{p}\sigma_p}^\dagger\prod^{N_\mathrm{op}}_{q=1}\hat{a}_{j_{q}\rho_q}
 \label{defmonomial}
\end{align}
where $g$ is a complex number and $\hat{g}$ is a single product of an arbitrary number ($N_\mathrm{op}$) of creation and annihilation operators
that create/annihilate orbitals with spatial orbital indices labelled by $i$, $j$ and $S_z$ indices labelled by $\sigma$, $\rho$.
We refer to this Hamiltonian as the excitation Hamiltonian because it only involves a single ``excitation'' term $\hat{g}$ and its Hermitian conjugate.
To specify $\hat{g}$ the FQE can digest a \texttt{FermionOperator} from OpenFermion.
If $\hat{H}_{\text{excite}}$ is diagonal, i.e. a polynomial of number operators (such as $\hat{a}^\dagger_{1\alpha} \hat{a}^\dagger_{2\alpha} \hat{a}_{1\alpha} \hat{a}_{2\alpha} = - \hat{n}_{1\alpha} \hat{n}_{2\alpha}$, with $\hat{n}_{1\alpha} =\hat{a}^\dag_{1\alpha}a_{1\alpha}$ and similarly for $\hat{n}_{2\alpha}$) evolution of the wave function can be performed using the techniques described later in Sec.~\ref{section:diagonal}.
When $\hat{H}_{\text{excite}}$ is not diagonal and contains no repeated indices (such as $\hat{a}^\dagger_{4\alpha} \hat{a}^\dagger_{2\beta} \hat{a}_{1\alpha} \hat{a}_{3\beta}$), 
evolution of the wavefunction is accomplished by 
\begin{align}\label{eq:monomial_evolv}
e^{-i(\hat{g} + \hat{g}^{\dagger})\epsilon } = 1 &+ \left[\cos(\epsilon|g|)-1 -i \hat{g}^\dagger\frac{\sin(\epsilon|g|)}{|g|} \right]\hat{\mathcal{P}}_{\hat{g}\hat{g}^\dagger}\\\nonumber
&+ \left[\cos(\epsilon|g|)-1 -i \hat{g}\frac{\sin(\epsilon|g|)}{|g|} \right]\hat{\mathcal{P}}_{\hat{g}^\dagger\hat{g}} 
\end{align}
where we define $\hat{\mathcal{P}}_{x}$ as the projector onto the basis of determinants that are not annihilated by $\hat{x}$. For other recent applications of this relation in quantum chemistry see e.g. Refs.~\cite{Filip2020JCP},~\cite{Evangelista2019JCP}, and Ref.~\cite{Chen2021JCTC} in the context of quantum and quantum-inspired algorithms.  A  derivation of Eq.~\eqref{eq:monomial_evolv} is presented in  Appendix~\ref{app:monomial_evolution}. When $\hat{H}_{\text{excite}}$ is not diagonal but has several repeated indices as well (such as $\hat{a}^\dagger_{1\alpha} \hat{a}^\dagger_{2\beta} \hat{a}_{1\alpha} \hat{a}_{3\beta} = - \hat{n}_{1\alpha} \hat{a}^\dagger_{2\beta} \hat{a}_{3\beta}$), a hybrid approach is used with the only additional complication being the fermion parity evaluation that arises from operator reordering.

Time evolution of a FQE wavefunction is implemented as a method of the \texttt{Wavefunction} object and can be easily accessed through the \texttt{Wavefunction} object interface. For example, the code snippet
\begin{lstlisting}[language=Python]
from openfermion import FermionOperator
import fqe

wf = fqe.Wavefunction([[4, 0, 6]])
wf.set_wfn(strategy='random')
i, j, theta = 0, 1, 2 / 3
op = (FermionOperator(((2 * i, 1), (2 * j, 0)), coefficient=-1j * theta) + 
      FermionOperator(((2 * j, 1), (2 * i, 0)), coefficient=1j * theta))
new_wfn = fqe_wfn.time_evolve(1.0, op)
\end{lstlisting}
performs evolution of a random state $\psi_{0}$ with a one-particle excitation Hamiltonian acting on the $\alpha$-spin sector
\begin{align}
\vert \psi_{f}\rangle = e^{-i \theta (\hat{a}_{i\alpha}^{\dagger}\hat{a}_{j\alpha} + \hat{a}_{j\alpha}^{\dagger}\hat{a}_{i\alpha})}\vert \psi_{0}\rangle.
\end{align}
The \texttt{Wavefunction} object correctly evolves this type of sparse Hamiltonian so long as the Hamiltonian commutes with the underlying wavefunction symmetries. However, the user is responsible for providing a Hamiltonian of this form, and the FQE does not correctly handle the evolution of Hamiltonians which break the specified wavefunction symmetry.

Evolving under a single fermionic excitation Hamiltonian is the basis for arbitrary fermionic quantum circuit evolution and can be used to simulate arbitrary fermionic Hamiltonian evolution. Because of the isomorphism between fermion and qubit spaces, general qubit Hamiltonians, so long as they are symmetry preserving in the same sense, can be simulated with little overhead within the same scheme either by modifying the code to ignore signs arising from fermion parity or by explicitly inserting swap operations. This raises the possibility to simulate large quantum circuits
associated with non-fermionic Hamiltonian evolution, that benefit from the memory saving associated with simulating the fixed ``particle'' or ``spin'' sectors of the corresponding qubit Hilbert space. 

Though evolution by single excitation Hamiltonian provides a  primitive to implement 
many fermionic circuit simulations, the FQE also implements efficient time-evolution routines for Hamiltonians with special forms and for generic quantum chemical Hamiltonians.  These will be discussed in more detail in the following sections.

\subsection{Evolution of Dense Hamiltonians}
The FQE provides special routines to evolve sums of excitation Hamiltonians through series expansions. We begin by discussing dense Hamiltonians, which for the purposes of this work we define as a weighted sum over all possible excitation Hamiltonians with the same index structure. 
A dense two-particle Hamiltonian, for which the FQE implements special routines, would take the form
\begin{align}
\hat{H} = \sum_{ijkl}\sum_{\sigma\sigma'\rho\rho'} V_{i\sigma,j\rho,k\sigma',l\rho'}\hat{a}_{i\sigma}^{\dagger}\hat{a}_{j\rho}^{\dagger}\hat{a}_{k\sigma'}\hat{a}_{l\rho'}
\label{Hdense}
\end{align}
where $\hat{H}$ is Hermitian. 
In addition, specialized code is implemented for spin-conserving spin-orbital Hamiltonians ($\sigma = \sigma'$ and  $\rho = \rho'$ in Eq.~\eqref{Hdense}),
\begin{align}
    \hat{H} = \sum_{ijkl} \sum_{\sigma\rho} V_{i\sigma, j\rho, k\sigma, l\rho} \hat{a}_{i\sigma}^{\dagger}\hat{a}_{j\rho}^{\dagger}\hat{a}_{k\sigma}\hat{a}_{l\rho}. \label{Hsso}
\end{align}
and for spin-free Hamiltonians ($V_{ijkl} \equiv V_{i\sigma,j\rho,k\sigma',l\rho'}$ in Eq.~\eqref{Hsso}),
\begin{align}
\hat{H}_\mathrm{sf} = \sum_{ijkl} \sum_{\sigma\rho} V_{ijkl} \hat{a}_{i\sigma}^{\dagger}\hat{a}_{j\rho}^{\dagger}\hat{a}_{k\sigma}\hat{a}_{l\rho}, \label{Hsp}
\end{align}
which frequently appear in molecular and materials systems. Other cases which have specialized subroutines discussed below include sparse Hamiltonians (arbitrary sums of excitation Hamiltonians), quadratic Hamiltonians, and diagonal pair Hamiltonians.

Time evolution with such dense Hamiltonians is performed by means of series expansions such as the Taylor and Chebyshev expansions.
When the time step $t$ is taken to be small, the operator exponential can be efficiently computed by a Taylor expansion
that converges rapidly,
\begin{align}
    \exp(-i\hat{H} t) |\Psi\rangle = \sum_n \frac{1}{n!} (-i\hat{H}t)^n  |\Psi\rangle.
\end{align}
We evaluate this by recursively computing the action of the operator on the wave functions,
\begin{align}
    |\Psi_{n+1}\rangle = \hat{H} |\Psi_n\rangle \label{app1}
\end{align}
with $|\Psi_0\rangle \equiv |\Psi\rangle$, using which the previous expression can be rewritten as 
\begin{align}
    \exp(-i\hat{H} t) |\Psi\rangle =  \sum_n \frac{(-it)^n}{n!} |\Psi_n\rangle
\end{align}
We evaluate the norm of each term, $|t^n|||\Psi_n\rangle|/n!$, and when the norm becomes smaller than the given threshold,
we stop the computation. Alternatively, one can specify the number of terms in the expansion, or both.

When the spectral range of the operator (i.e., extremal eigenvalues $[\epsilon_\mathrm{min}, \epsilon_\mathrm{max}]$) is known or can be estimated in advance, 
the expansion can be made more robust by using the Chebyshev expansion \cite{Kosloff1994ARPC}. The Chebyshev expansion is known to be
a near optimal approximation of the exponential in a minimax sense. First, we scale the operator such that the eigenvalues are in $[-1, +1]$,
\begin{align}
    \hat{H}' = \frac{\hat{H} + \epsilon_\mathrm{shift}}{\Delta \epsilon}
\end{align}
where
$ \Delta \epsilon= (\epsilon_\mathrm{max} - \epsilon_\mathrm{min})/2w'$ and
$ \epsilon_\mathrm{shift} = - (\epsilon_{\max} + \epsilon_{\min})/2$.
The factor $w'$ is a number that is slightly smaller than 1 to ensure that the eigenvalues 
do not lie outside the window due to numerical noise or insufficient accuracy in the estimation of 
extremal eigenvalues. In the FQE, we use $w' = 0.9875$. Then we expand as
\begin{align}
    \exp(-i\hat{H} t)  |\Psi\rangle &=  \exp(i\epsilon_\mathrm{shift} t) \exp(-i \Delta \epsilon \hat{H}' t)|\Psi\rangle\nonumber\\
    &= \exp(i\epsilon_\mathrm{shift} t)  \sum_n 2 J_n(\Delta \epsilon t)(-i)^n  |\tilde{\Psi}_n\rangle
\end{align}
Here $J_n(x)$ is the modified Bessel function of the first kind, and $|\tilde{\Psi}_n\rangle$ is obtained
by recursion as
\begin{align}
   |\tilde{\Psi}_{n+1}\rangle = 2 \hat{H}' |\tilde{\Psi}_n\rangle  - |\tilde{\Psi}_{n-1}\rangle \label{app2}
\end{align}
We stop the expansion when the contribution from rank $n$ becomes smaller than the given threshold, as in the Taylor expansion above.

The dense evolution routines can be accessed through the \texttt{Wavefunction} interface which intelligently dispatches to the specialized routines depending on the form of the Hamiltonian the user provides. As an example, below is a code snippet assuming the user has defined an OpenFermion \texttt{MolecularData} object called `molecule'.  The Wavefunction interface provides access to the \texttt{apply$\_$generated$\_$unitary} routine which has options for each series expansion evolution described above.
\begin{lstlisting}[language=Python]
from fqe.openfermion_utils import integrals_to_fqe_restricted

dt = 0.23  # evolution time
oei, tei = molecule.get_integrals()
# A dense fqe.RestrictedHamiltonian object
fqe_rham = integrals_to_fqe_restricted(oei, tei)

wfn = fqe.Wavefunction([[molecule.n_electrons, 0, molecule.n_orbitals]])
wfn.set_wfn(strategy='hartree-fock')
new_wfn1 = wfn.time_evolve(dt, fqe_rham) 

# equivalent path
# Taylor expansion algorithm
new_wfn2 = wfn.apply_generated_unitary(dt, 'taylor', fqe_rham) 
\end{lstlisting}

The above algorithms that use the series expansion are also employed for time evolution with
sparse Hamiltonians, which are arbitrary weighted sums of multiple excitation Hamiltonians.
In this case, the \texttt{apply} function used in Eqs.~\ref{app1} and \ref{app2} is overloaded by a special function 
for sparse Hamiltonians, in which the action of the operator is evaluated one term at a time.

\subsection{Special routines for other structured Hamiltonians}
\subsubsection{Diagonal pair Hamiltonians\label{section:diagonal}}
\begin{figure}[H]
    \centering
    \includegraphics[width=7.0cm]{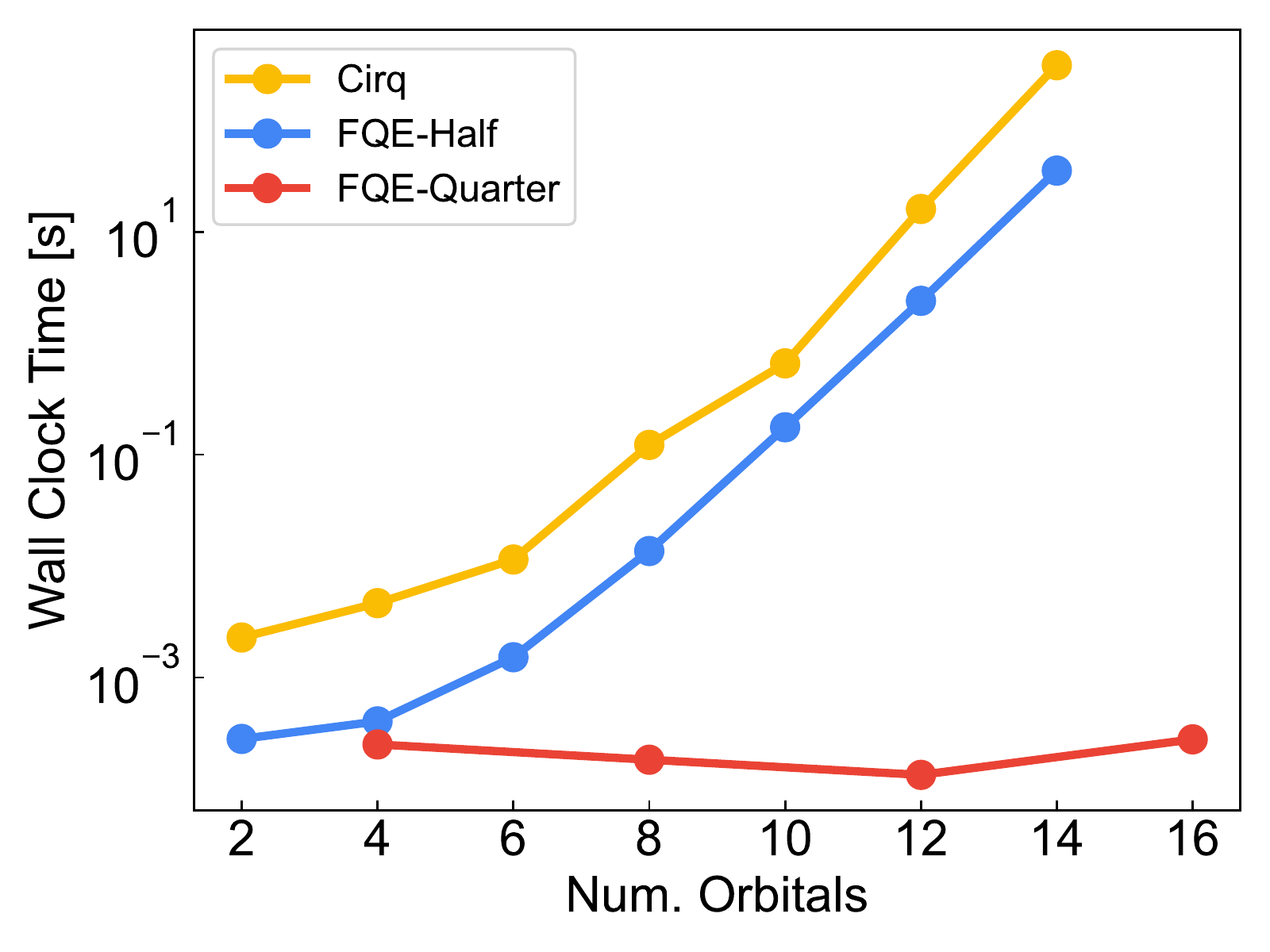}
    \includegraphics[width=7.0cm]{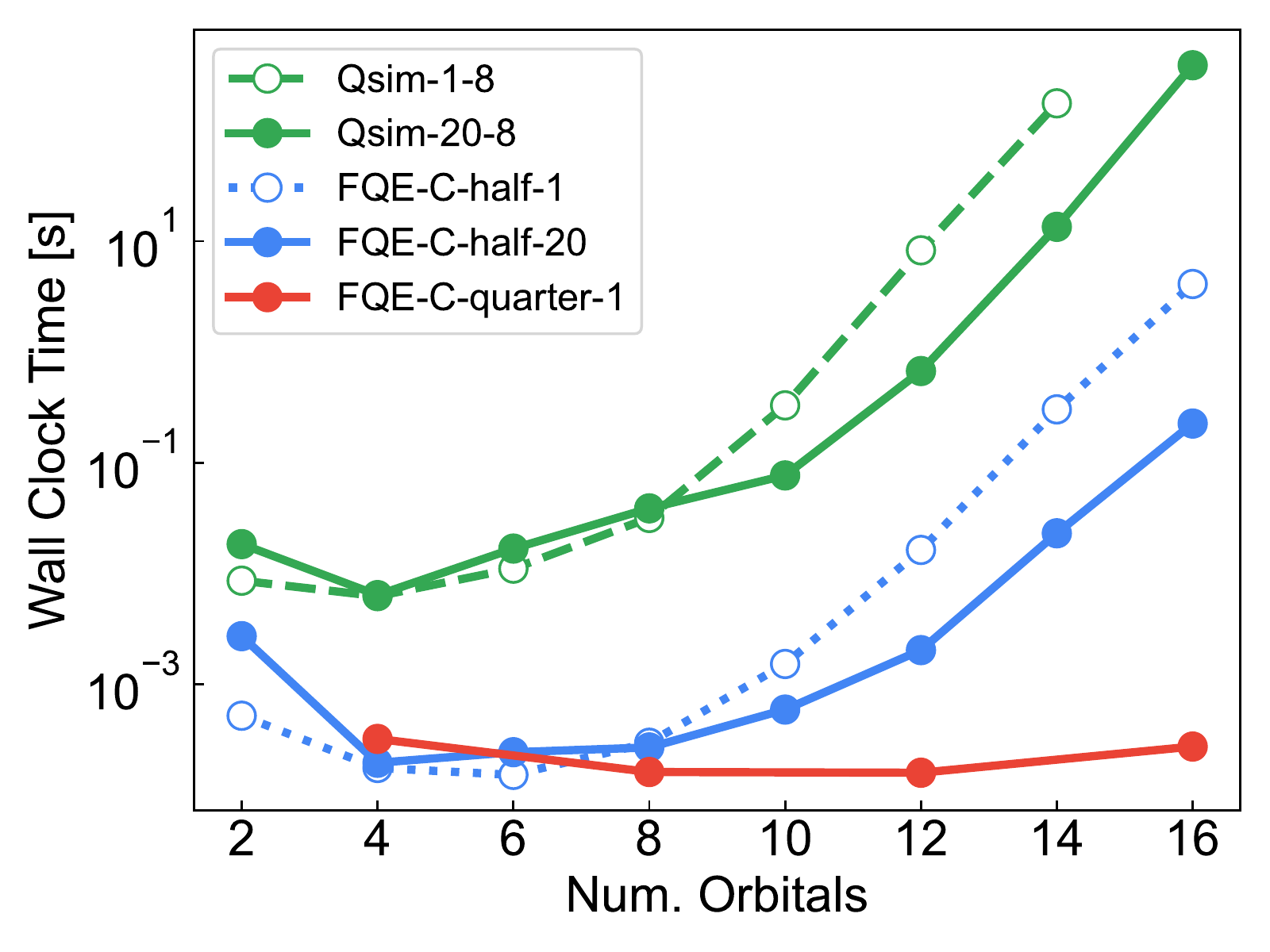}
    \caption{Run time comparisons for implementing time evolution of a random diagonal Hamiltonian: \textit{left} The Python reference implementation of FQE is compared against Cirq's Python circuit evolution routines; \textit{right}  C-accelerated FQE is compared against Qsim, a highly optimized C++ circuit simulator with and without threading. FQE computes the results in-place. Qsim is executed in single precision mode with one (green dashed line) and twenty (green solid line) threads with gate gate-fusion set to eight through the \texttt{qsimcirq} python interface. Quantum circuits for the time evolution generated by the diagonal Coulomb Hamiltonian are constructed using $4 M^{2}$ \textsc{CPHASE} gates where $M$ is the number of spatial orbitals.  FQE at half filling with one and twenty threads is shown in blue while quarter (single thread) filling is in red. }
    \label{fig:diag_coulomb}
\end{figure}

Hamiltonians that are diagonal in the determinant basis are particularly simple as they are polynomials of number operators with terms that all commute. 
Evolution under such Hamiltonians appears as an important quantum circuit for chemical Hamiltonian Trotter steps and certain variational ans\"atze~\cite{lee2018generalized, motta2018low}.  The FQE leverages the mutual commuting nature of the terms of Hamiltonians of this form to efficiently evolve under the generated unitaries. 
Specifically,  
the action of the unitary generated by diagonal pair Hamiltonians of the form
\begin{align}
\label{eq:diagH}
\hat{H}_{\text{diag}} = \sum_{rs} W_{rs}\hat{n}_{r}\hat{n}_{s},
\end{align}
where $\hat{n}_{r}$ is the spin-summed number operator for spatial orbital $r$ (i.e., $\hat{n}_{r} = \sum_\sigma \hat{a}^\dagger_{r\sigma} \hat{a}_{r\sigma}$),
is constructed by iterating over all determinants and generating a phase associated with the coefficient $W_{rs}$ for each determinant. A direct wall clock time comparison to other simulators is made in Figure~\ref{fig:diag_coulomb} by translating evolution under a random diagonal pair  Hamiltonian (of the form in Eq.~\eqref{eq:diagH}) into a series  of \textsc{CPHASE} gates with phases corresponding to $2W_{rs}$.  For the half filling circuits on 14 orbitals (i.e., 14 electrons in 28 spin orbitals to be
represented by 28 qubits) we observe that the python reference implementation of FQE is approximately eight times faster than Cirq and five times faster than single-threaded Qsim. The C-accelerated FQE single threaded is five hundred times faster than Qsim single threaded with default gate fusion. The C-accelerated FQE on twenty threads is twenty times faster than QSim using twenty threads and gate-fusion level eight for the same 14 orbital half filled system.  FQE performance gains for lower filling fractions are even more substantial.

\subsubsection{Quadratic Hamiltonians}\label{sec:quad_evolve}
Evolving states by quadratic Hamiltonians is another important algorithmic primitive. In the quantum circuit context, it is closely related to matchgate circuit simulation~\cite{PhysRevA.65.032325}.
The FQE performs this task by transforming the wave functions
into the orbital representation that diagonalizes the quadratic Hamiltonian, evolving in the quadratic Hamiltonian eigenbasis, and then returning to the original basis.  The algorithm is based on those in the quantum chemistry literature \cite{Malmqvist1986IJQC, Atchity1999JCP, Mitrushchenkov2007MP}
that utilize the LU decomposition, with an improvement in the handling of pivoting in the LU decomposition.
Using the unitary $\mathbf{X}$ that diagonalizes 
the operator matrix in $\hat{A} = \sum_{ij}(\mathbf{A})_{ij}\hat{a}^\dagger_i \hat{a}_j$, i.e., $\mathbf{X}^\dagger \mathbf{A}\mathbf{X} = \mathbf{a}$,
one obtains
\begin{align}
     \exp(-i\hat{A}t) |\Psi\rangle &= \hat{T}(\mathbf{X}^\dagger) \exp(-i\hat{a}t) \hat{T}(\mathbf{X}) |\Psi\rangle
\end{align}
where $\hat{a}$ is the diagonal operator after orbital rotation, and 
$\hat{T}(\mathbf{X})$ is the transformation on the wave function due to the change in the orbital basis described by
the unitary $\mathbf{X}$. Here we ignore the pivoting for brevity;
see Appendix \ref{section:quad} for more algorithmic details.
The overall cost of rotating the wave function (i.e., action of $\hat{T}(\mathbf{X})$) is roughly equivalent to the cost
to apply a single dense one-particle linear operator to a wave function.
The diagonal operator is then applied to the rotated wave function, followed by the back transformation to the wave function
in the original orbital basis.

\begin{figure}[H]
    \centering
    \includegraphics[width=7.0cm]{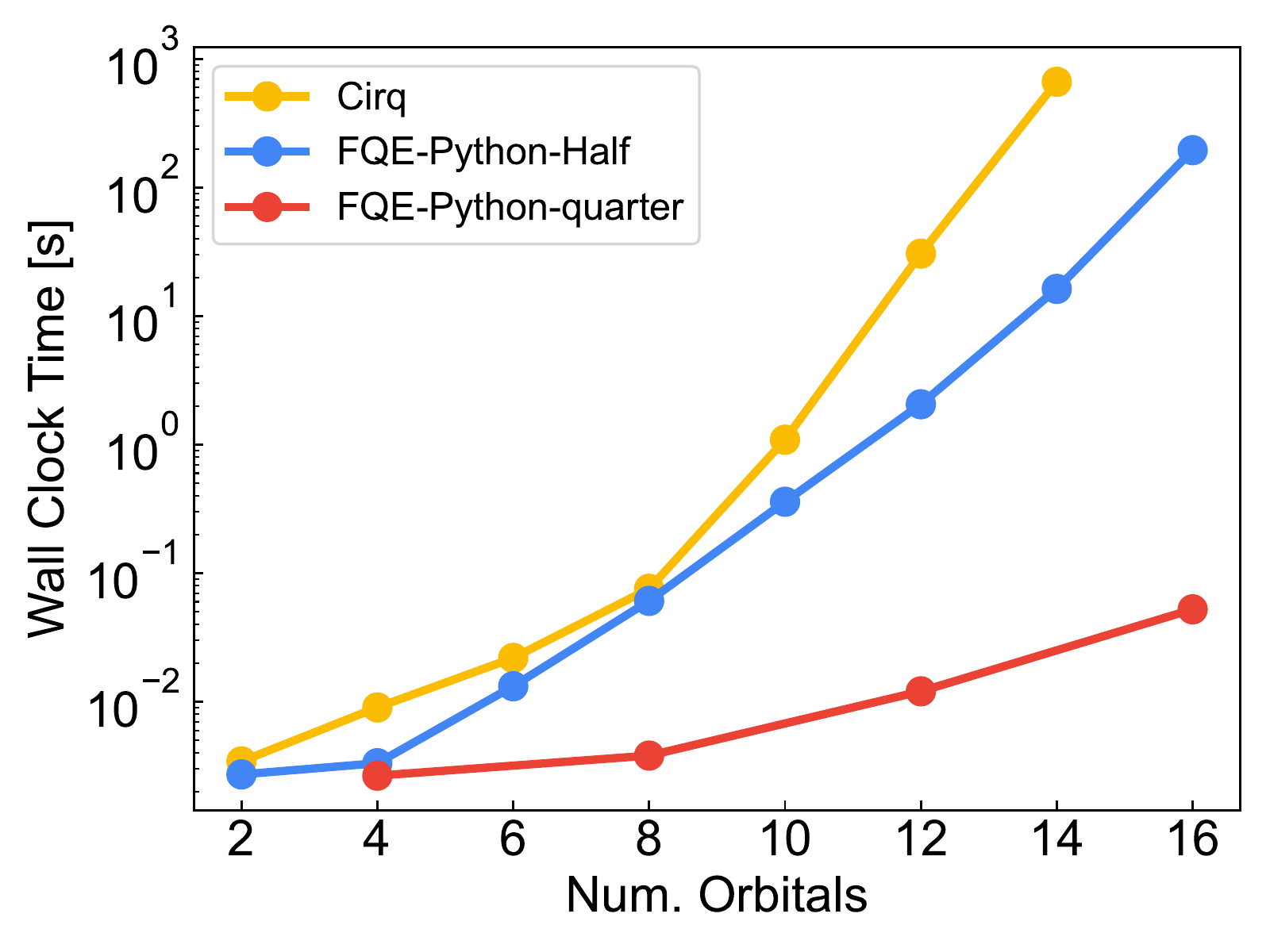}
    \includegraphics[width=7.0cm]{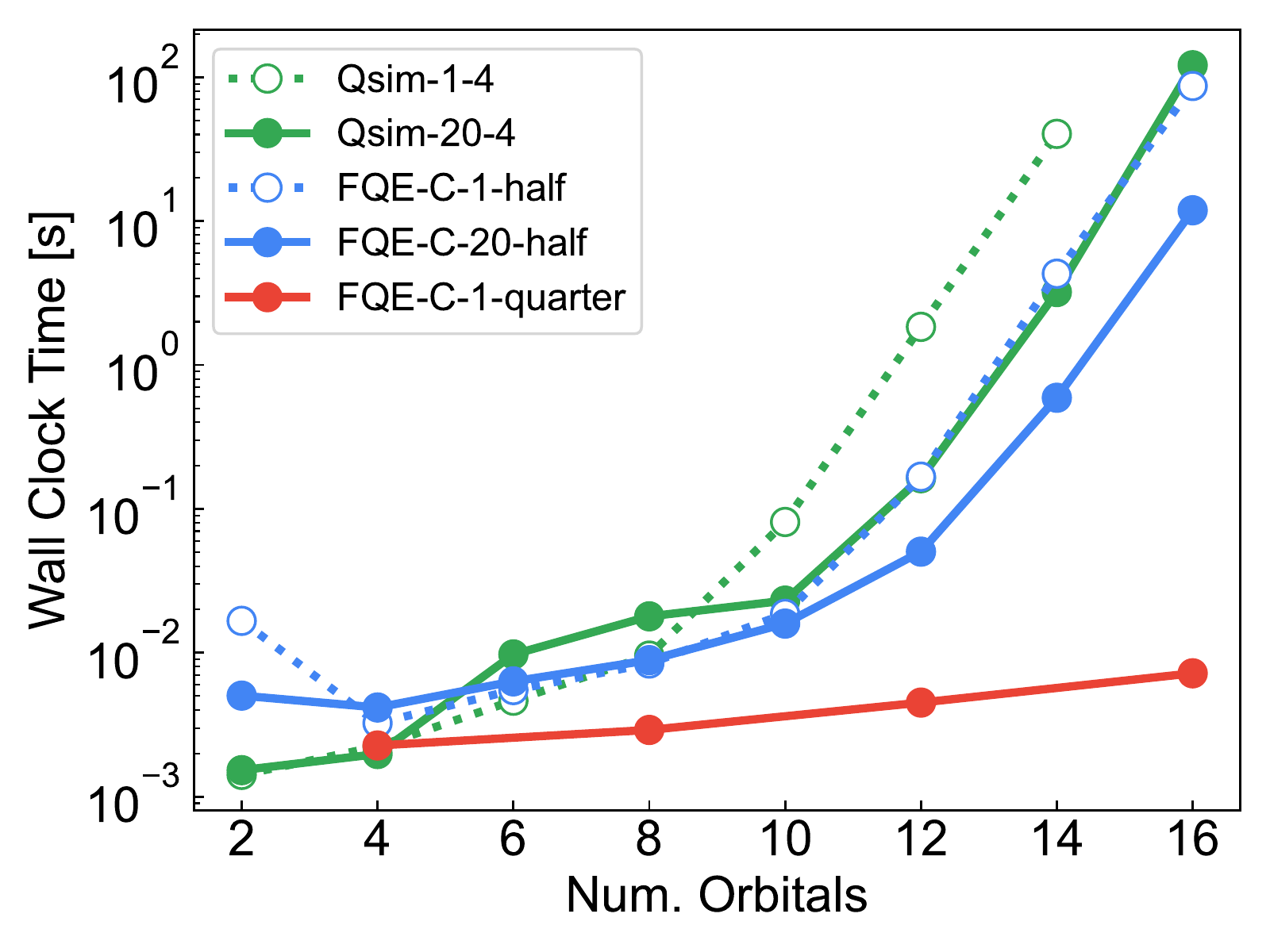}
    \caption{Wall clock time comparison for a random quadratic Hamiltonian evolution at half filling and quarter filling, \textit{left} between Cirq and the Python implementation in FQE and \textit{right} between Qsim and C-accelerated FQE. Note that Cirq and Qsim run times are unaffected by filling fraction.  Qsim is executed accessed through the \texttt{qsimcirq} Python interface.  Each Qsim trace is labeled by how many threads are used (one or twenty) and the gate-fusion level (four).  Quantum circuits for the time evolution generated by the quadratic Hamiltonian are constructed using $ 2M^{2}$ $\exp{(-i \theta (XY-YX)/2})$ gates where $M$ is the number of spatial orbitals. FQE evolves the quadratic Hamiltonian in Section~\ref{sec:quad_evolve}. }
    \label{fig:quadratic_time}
\end{figure}

Figure~\ref{fig:quadratic_time} compares the wall clock time required to evolve a random quadratic Hamiltonian.  For Cirq and Qsim, a circuit is generated using the optimal Givens rotation decomposition construction of Clements~\cite{clements2016optimal} in OpenFermion which is translated into a sequence of $2M^{2}$ gates of the form $\exp{(-i \theta (XY-YX)/2)}$ and $\mathrm{Rz}$. 
For the half filling circuits on fourteen orbitals (twenty eight qubits) we observe that the python-FQE LU decomposition algorithm is approximately forty times faster than Cirq.
 The single-thread C-accelerated FQE is approximately ten times faster than single-threaded Qsim with gate-fusion set to four.  For twenty threads the C-accelerated FQE is five times faster than Qsim.  Lower filling fractions show even more substantial savings. This illustrates the large advantage physics, symmetry, and algorithmic considerations can provide when targeting specialized quantum circuit emulation. 

\section{Operator Action on the Wavefunction}\label{sec:op_action}
The FQE provides a variety of methods to apply fermionic operators to states. 
The general code allows a user to define an arbitrary individual excitation operator and apply it to a wavefunction.
This functionality constructs the action of the user defined operator by computing the sparse matrix elements of such an operator in the determinant basis, which are then contracted 
against the wavefunction coefficients to obtain the action of the operator on the state. 

The FQE also provides efficient methods to apply dense fermionic operators (such as a dense Hamiltonian) to wave functions, because these are 
the fundamental building blocks for fermionic time evolution under  dense Hamiltonians (see, for example, Eqs.~\eqref{app1} and \eqref{app2}).
It is equally important for reduced density matrix (RDM) construction, efficient access to which is
a crucial aspect of many fermionic simulation algorithms.
In the following, we present the algorithms  to apply dense spin conserving, spin-free Hamiltonians (Eq.~\eqref{Hsp}), but the FQE implements similar algorithms for 
$S_z$ conserving spin-orbital Hamiltonians (Eq.~\eqref{Hsso}) and $S_z$ non-conserving Hamiltonians (Eq.~\eqref{Hdense}) as well.

\subsection{Knowles--Handy algorithm}
In the Knowles--Handy algorithm~\cite{Knowes1984CPL}, a resolution of the identity (RI) is inserted in the $n$-electron determinant space after operator reordering
($\hat{1} = \sum_K |K\rangle \langle K|$) as 
\begin{align}
\langle I|\hat{H}_\mathrm{sf} |\Psi\rangle &= - \sum_{\sigma \rho}\sum_{ik,K} \langle I | \hat{a}^\dagger_{i\sigma} \hat{a}_{k\sigma} |K\rangle 
\left(\sum_{jl,I} V_{ijkl} \langle K|\hat{a}^\dagger_{j\rho} \hat{a}_{l\rho}|J\rangle C_J\right) \nonumber\\
&+ \sum_\sigma \sum_{il} \left(\sum_k V_{ikkl}\right) \langle I| \hat{a}^\dagger_{i\sigma} \hat{a}_{l\sigma}|J\rangle C_J,
\end{align}
where 
 $I$, $J$, and $K$ label determinants and
$C_I$ is the wave function coefficient associated with determinant $|I\rangle$. 
The second term is trivially evaluated. The primitive operations for the first term are 
\begin{align} 
D^{J}_{ij} &= \sum_I\sum_\sigma \langle J|\hat{a}^\dagger_{i\sigma} \hat{a}_{j\sigma}|I\rangle C_{I}, \label{kh1}\\
\tilde{D}^{J}_{ij} &= \sum_{kl} V_{ikjl} D^{J}_{kl}, \label{kh2}\\
R^{I} &= \sum_J\sum_\sigma \langle I|\hat{a}^\dagger_{i\sigma} \hat{a}_{j\sigma}|J\rangle \tilde{D}^{J}_{ij}. \label{kh3}
\end{align}
This form of the RI insertion has the advantage that it can be easily generalized to apply  3- and 4-particle operators, since these primitive operations can be performed recursively, i.e., 
\begin{align}\label{eq:intermediate_tensors}
    E^{J}_{ij,kl}= \sum_I\sum_\sigma \langle J|\hat{a}^\dagger_{i\sigma} \hat{a}_{j\sigma}|I\rangle D^I_{kl},\\
    D^{I}_{ij}= \sum_J \sum_\sigma \langle I|\hat{a}^\dagger_{k\sigma} \hat{a}_{l\sigma}|J\rangle E^{J}_{ij,kl}.
\end{align}
The step in Eq.~\eqref{kh2} is typically a computational hot spot and is performed using efficient \texttt{numpy} functions.

For efficiency, the expectation values 
$\langle I_\sigma\vert \hat{a}_{i\sigma}^{\dagger}\hat{a}_{j\sigma}\vert J_\sigma\rangle$
are precomputed for the $\alpha$- and $\beta$-strings and stored in an FQE object \texttt{FciGraph}.
To facilitate the evaluation, the intermediate tensors are also stored using the $\alpha$- and $\beta$-strings; for example,
$D^I_{ij}$ in Eq.~\eqref{kh1} is stored as a four-index tensor $D^{I_{\alpha}I_{\beta}}_{ij}$ where $I = I_\alpha\otimes I_\beta$.
When the Hamiltonian breaks spin symmetry (and, hence, $S_z$ symmetry), the mappings between strings that have $N_\sigma \pm 1$ electrons are also generated
($\langle I'_\sigma\vert \hat{a}_{i\sigma}^{\dagger}\vert J_\sigma\rangle$ and 
$\langle I''_\sigma\vert \hat{a}_{i\sigma}\vert J_\sigma\rangle$)
which are stored in the FQE object \texttt{FciGraphSet}. From these quantities, $\langle I | \hat{a}_{i\sigma}^\dagger \hat{a}_{j\sigma} |J\rangle$ can be  easily computed on the fly. 
\subsection{Harrison--Zarrabian algorithm for low filling}
Low filling cases are computed using the Harrison--Zarrabian algorithm \cite{Harrison1989CPL}, which is closely related to
the algorithm developed by Olsen \textit{et al.} \cite{Olsen1988JCP}. This algorithm inserts the RI in the 
$n-2$ electron space ($\hat{1} = \sum_L |L\rangle \langle L|$) as  
\begin{align}
    \langle I|\hat{H}_\mathrm{sf}|\Psi\rangle = \sum_{\sigma \rho}\sum_{ij,L} \langle I | \hat{a}^\dagger_{i\sigma} \hat{a}^\dagger_{j\rho} |L\rangle  \left(\sum_{kl,J} V_{ijkl} \langle L|\hat{a}_{k\sigma} \hat{a}_{l\rho}|J\rangle C_J\right),
\end{align}
where $I$ and $J$ label a determinant with $n$ electrons and $L$ labels determinants with $n-2$ electrons.
This is computed by the following primitive steps,
\begin{align}
F^{L}_{ij,\sigma\rho} &= \sum_I \langle L|\hat{a}_{k\sigma} \hat{a}_{l\rho}|I\rangle C_{I}, \label{hz1}\\
\tilde{F}^{L}_{ij,\sigma\rho} &= \sum_{kl} V_{ijkl} F^{L}_{kl,\sigma\rho}, \label{hz2}\\
R^{I} &= \sum_L\sum_{\sigma\rho} \langle I|\hat{a}^\dagger_{i\sigma} \hat{a}^\dagger_{j\rho}|L\rangle \tilde{F}^{L}_{ij,\sigma\rho}. \label{hz3}
\end{align}
The advantage of this algorithm, in comparison to the Knowles--Handy algorithm, is that
the number of determinants for the RI can be far smaller than the number of the original determinants.
This is pronounced when the wave function is at low filling.
The disadvantage, however, is that there is certain overhead in computational cost because one cannot perform
 spin summation in Eq.~\eqref{hz1}.
Therefore, the FQE switches to this algorithm
when the filling is smaller than 0.3. Note that a similar algorithm can be devised for the high-filling cases
by sorting the operators in the opposite order, but this is not implemented in the FQE.
When such a system is considered, one can reformulate the high-filling problem into
a low-filling one by rewriting the problem in terms of the hole operators $\hat{b}_{i\sigma} = 1 - \hat{a}_{i\sigma}$.

\subsection{Olsen's algorithm}
The approach of Knowles and Handy, while being simple to implement with optimized linear algebra routines in Python, can be further improved by taking advantage of the sparsity in the $D^J_{ij}$ appearing in Equation~\ref{kh1}. For example, $D^J_{ij}$ will be zero if determinant $J$ has orbital $i$ unoccupied or orbital $j$ doubly occupied. This means that extra work is performed during the most expensive step of the algorithm, Equation~\ref{kh2}. Olsen and co-workers were able to leverage this sparsity by avoiding explicit reference to intermediate states~\cite{Olsen1988JCP}. In this approach, the same replacement lists used by Knowles and Handy in the construction of the $D^J_{ij}$ are used to loop over only the non-zero Hamiltonian elements connecting bra and ket. The essence of the algorithm is to structure the loops in such a way that vector operations can be used. Details of the algorithm are given by Olsen {\it et al.}~\cite{Olsen1988JCP} and clearly reviewed by Sherrill and Schaefer~\cite{Sherrill1999}. While Olsen's algorithm cannot be efficiently implemented in pure Python, it is the basis of the FQE C extension for applying dense, two-particle Hamiltonians.

\subsection{RDM computation}
The FQE provides efficient routines using the intermediate tensors in the above algorithms to efficiently compute $1$- through 4-RDMs.
For example, a spin-summed 2-RDM element $\Gamma_{ijkl}$ defined as 
\begin{align}
\Gamma_{ijkl}=  \sum_{\sigma\rho}\langle \Psi| \hat{a}^\dagger_{i\sigma} \hat{a}^\dagger_{j\rho}\hat{a}_{k\sigma} \hat{a}_{l\rho} |\Psi\rangle   \label{eq:spinsumdm}
\end{align}
can be computed either using the intermediate tensor in the Knowles--Handy algorithm (Eq.~\eqref{kh1}),
\begin{align}
\Gamma_{ijkl}= - \sum_{I} D^{I\ast}_{ki} D^I_{jl} + \delta_{jk} \sum_I D^{I\ast}_{li} C_I,
\end{align}
or, in low filling cases, using the intermediate in the Harrison--Zarrabian algorithm,
\begin{align}
\Gamma_{ijkl} = - \sum_{\sigma\rho} \sum_{L} F^{L\ast}_{ij,\sigma\rho} F^L_{kl,\sigma\rho}.
\end{align}
The corresponding generalization of Olsen's algorithm is used when the C extensions are enabled.
In both Python algorithms, efficient \texttt{numpy} functions for matrix--matrix and matrix--vector multiplication are used.
Spin-orbital RDMs (i.e. without the spin summation in Eq.~\eqref{eq:spinsumdm}) are computed similarly.
In addition to the standard (i.e., particle) RDMs, the FQE can compute RDMs that correspond to other operator orders, e.g.,
hole RDMs. This is done by performing Wick's theorem as implemented in \texttt{fqe.wick} and evaluating the resulting
expression using the particle RDMs.
For example, the spin-summed 2-body hole RDM element ($\Gamma^h_{ijkl}$) defined as 
\begin{align}
    \Gamma^h_{ijkl} =\sum_{\sigma\rho}\langle \Psi| \hat{a}_{i\sigma} \hat{a}_{j\rho}\hat{a}^\dagger_{k\sigma} \hat{a}^\dagger_{l\rho} |\Psi\rangle 
\end{align}
is computed as
\begin{align}
\Gamma^h_{ijkl} = \Gamma_{klij} + \delta_{jk}\Gamma_{il} - 2\delta_{ik}\Gamma_{jl} - 2\delta_{jl}\Gamma_{ik}.
\end{align}
The FQE implements the spin-orbital counterpart of this feature as well.

For spin conserving Hamiltonians, the FQE also provides up to $4$-body spin-summed RDMs and up to $3$-body spin-orbital RDMs in the OpenFermion format. An example is:
\begin{lstlisting}[language=Python]
wf = fqe.Wavefunction([[4, 0, 6]])
wf.set_wfn(strategy='random')

spin_sum_opdm = wf.expectationValue('i^ j')
spin_sum_oqdm = wf.expectationValue('i j^')
spin_sum_tpdm = wf.expectationValue('i^ j^ k l')

opdm, tpdm = wf.sector((4, 0)).get_openfermion_rdms()
d3 = wf.sector((4, 0)).get_d3()  # returns spinful 3-RDM in openfermion ordering
\end{lstlisting}
The implementation to compute RDMs offers a significant improvement over OpenFermion's native RDM generators which map Pauli operator expectations to RDM elements.  As an example of this efficient RDM computation the FQE library provides a base implementation of the energy gradient due to a two-particle generator 
$\hat{G}_{i\sigma, j\rho, k\sigma', l\rho'}=  \hat{a}^\dagger_{i\sigma} \hat{a}^\dagger_{j\rho} \hat{a}_{k\sigma'} \hat{a}_{l\rho'}$, which corresponds to evaluating
$g_{i\sigma, j\rho, k\sigma', l\rho'} = \langle \Psi \vert [\hat{H}, \hat{G}_{i\sigma, j\rho, k\sigma', l\rho'}]\vert \Psi \rangle$ for all spatial and spin indices.

\section{Closing thoughts and future directions}
The Fermionic Quantum Emulator (FQE) described above is an open source library under the Apache 2.0 license~\cite{fqe_2020}. Currently, the library is completely implemented in Python to facilitate extension and code reuse. Despite not being written in a high performance programming language, the FQE's algorithmic advantages allow us to outperform even heavily optimized quantum circuit simulators. In future releases, current computational bottlenecks will be addressed with additional performance improvements.
The Fermionic Quantum Emulator (FQE) provides a user-friendly and developer-friendly code stack without sacrificing performance. Algorithmic advantages allow the FQE to clearly outperform both Python-based and heavily optimized quantum circuit simulators, even at half filling, with our C-accelerated code. Away from half filling, the efficiency of FQE is even more pronounced. Extending the code to GPUs or other architectures may lead to even better performance.

Though this first release of the FQE focuses on exact evolution in a statevector representation, our long-term goal is to provide  implementations of fermionic circuit emulation that exploit symmetry within different simulation strategies. For example, approximate representation using fermionic matrix product states\sout{,} can also be adapted to this setting. 
Ultimately, a variety of such techniques will need to be explored to reach an honest assessment of possible quantum advantage when simulating molecular and materials systems.

\section{Acknowledgments}

NCR thanks Ryan LaRose for testing and feedback on the code.
\bibliographystyle{plainnat}
\bibliography{biblo,ryan_references_mirror}

\appendix
\section{Evolution under a Hermitian Hamiltonian generated by a sum of excitation operators}\label{app:monomial_evolution}
Here we discuss the unitary evolution associated with a fermionic excitation (Eq.~\eqref{defmonomial}) that is not diagonal and has no repeated orbital indices. In this case, one can trivially show that $\hat{g}\hat{g} = 0$ and $\hat{g}^\dagger\hat{g}^\dagger = 0$. Using these relations, it can be shown that
\begin{align}
    (\hat{g} + \hat{g}^\dagger)^n = \hat{g}\hat{g}^\dagger \hat{g} \cdots + \hat{g}^\dagger \hat{g} \hat{g}^\dagger \cdots.
\end{align}
In addition, $\hat{g}^{\dagger}\hat{g}$ and $\hat{g}\hat{g}^{\dagger}$ are both diagonal in the computational basis, because any of the unpaired operators in $\hat{g}$ would be paired with its conjugate in $\hat{g}^{\dagger}\hat{g}$ and $\hat{g}\hat{g}^{\dagger}$.
For example, let $\hat{g} = g \hat{a}^\dagger_4 \hat{a}^\dagger_2 \hat{a}_3 \hat{a}_1$ (we have omitted the spin indices for brevity), then
\begin{align}
    \hat{g}\hat{g}^\dagger &= |g|^2 \hat{a}^\dagger_4 \hat{a}^\dagger_2 \hat{a}_3 \hat{a}_1 \hat{a}_1^\dagger \hat{a}_3^\dagger \hat{a}_2 \hat{a}_4\nonumber\\
    &= |g|^2 \hat{n}_4 \hat{n}_2 (1-\hat{n}_3) (1-\hat{n}_1) 
\end{align}
It is worth noting that the parity associated with this reordering for normal-ordered $g$ (all creation operators to the left of annihilation operators) is always even.
We define the square root of this diagonal operator with the following phase convention,
\begin{align}
\sqrt{\hat{g}\hat{g}^\dagger} \equiv |g| \hat{n}_4 \hat{n}_2 (1-\hat{n}_3) (1-\hat{n}_1)\label{eq:sqrtgg}
\end{align}
where we used the fact that the action of the number operators gives 0 or 1, and therefore, they are idempotent.
Using these expressions, the Taylor expansion of the evolution operator is exactly re-summed
(where the summations are performed separately for the odd and even rank terms) as
\begin{align}
e^{-i(g + g^{\dagger})\epsilon } 
=& \sum_{n=0}^{\infty} \frac{(-i\epsilon)^n}{n!} (g + g^{\dagger})^n  \\\nonumber
=& -1 + \sum_{n=0}^{\infty} (-1)^{n} \frac{\left[(\hat{g}\hat{g}^\dagger)^n + (\hat{g}^\dagger \hat{g})^n\right] \epsilon^{2n}}{(2n)!}\\\nonumber
& -i \sum_{n=0}^{\infty} (-1)^{n} \frac{\left[ \hat{g}(\hat{g}^\dagger \hat{g})^n + \hat{g}^{\dagger}(\hat{g}\hat{g}^\dagger)^n\right] \epsilon^{2n + 1}}{(2n + 1)!} \\\nonumber
=& -1 + \cos(\epsilon\sqrt{\hat{g}\hat{g}^\dagger}) + \cos(\epsilon\sqrt{\hat{g}^\dagger \hat{g}})\\\nonumber
& - ig^\dagger \frac{1}{\sqrt{\hat{g}\hat{g}^\dagger}} \sin(\epsilon\sqrt{\hat{g}\hat{g}^\dagger}) 
-i \hat{g}\frac{1}{\sqrt{\hat{g}^\dagger \hat{g}}} \sin(\epsilon\sqrt{\hat{g}^\dagger \hat{g}}).
\end{align}
Denoting the projection to the set of determinants that are not annihilated by an operator $\hat{x}$ as  $\hat{\mathcal{P}}_{x}$, together with the convention in Eq.~\eqref{eq:sqrtgg},
this can be further simplified to Eq.~\eqref{eq:monomial_evolv}.

\section{Basis change: Evolution under quadratic Hamiltonians\label{section:quad}}
In this work, we make use of the following primitive. Given an orbital basis $\{ |\phi\rangle \}$ and many-electron wave function $|\Psi\rangle = \sum_{I(\phi)} C_{I(\phi)} |I(\phi)\rangle$ and a linear transformation $\hat{X} |\phi\rangle \to |\phi'\rangle$, we wish to re-express $|\Psi\rangle= \sum_{I(\phi')} C_{I(\phi')} |I(\phi')\rangle$ where $|I(\phi')\rangle$ is a determinant in the new orbital basis $\{ |\phi'\rangle\}$. There have been several discussions of how to implement this transformation efficiently \cite{Malmqvist1986IJQC, Atchity1999JCP, Mitrushchenkov2007MP}. In the
original work by Malmqvist \cite{Malmqvist1986IJQC}, it was understood that the transformation can be performed by the successive application of 
one-body operators to the wave functions, in which the operator was computed from the LU decomposition of the orbital transformation matrix $\mathbf{X}$ ($X_{ij} = \langle \phi_i|\phi_j'\rangle$).
Mitrushchenkov and Werner (MW) \cite{Mitrushchenkov2007MP}  later reported that the pivoting in the LU decomposition is necessary to 
make the transformation stable. In their work, a LAPACK~\cite{lapack} function was used to perform the LU decomposition
with pivoting on the rows of the orbital transformation matrix. This leads to reordering of the orbitals in the determinants and additional
phase factors in $C_{I(\phi')}$ that make it complicated to, for example, evaluate the overlap and expectation values.
Therefore, we have implemented a column-pivoted formulation of the MW algorithm, which is presented below. We will show the spin-free formulation (i.e. same transformation for $\alpha$ and $\beta$ orbitals)
for spin-conserving wave functions as an example,
but the procedure for the spin-broken case can be derived in the same way.

First, we obtain the column pivoted LU decomposition. This is done using \texttt{numpy.linalg.lu} (which pivots rows)
for the transpose of the orbital transformation matrix $\mathbf{X}$,
\begin{align}
    \mathbf{X}^T = \mathbf{\bar{P}}\mathbf{\bar{L}}\mathbf{\bar{U}},
\end{align}
where $\mathbf{\bar{L}}$ and $\mathbf{\bar{U}}$ are lower- and upper-triangular matrices, and the diagonal elements of $\mathbf{\bar{L}}$ are unit.
It is then easily seen that, by taking the transpose, one obtains
\begin{align}
    \mathbf{X} = \mathbf{\bar{U}}^T\mathbf{\bar{L}}^T\mathbf{\bar{P}}^T,
\end{align}
where $\mathbf{\bar{U}}^T$ and $\mathbf{\bar{L}}^T$ are lower and upper triangular, respectively. 
If one scales the rows and columns of $\mathbf{\bar{U}}^T$ and $\mathbf{\bar{L}}^T$, respectively, such that the diagonal elements of $\mathbf{\bar{U}}^T$ become unit, this can be rewritten as
\begin{align}
    \mathbf{X} = \mathbf{L}\mathbf{U}\mathbf{P}, \label{column}
\end{align}
in which $\mathbf{L}$ and $\mathbf{U}$ are lower- and upper-triangular matrices with the diagonal elements of $\mathbf{L}$ being unit, and
$\mathbf{P} = \mathbf{\bar{P}}^T$.
These are done in the \texttt{ludecomp} and \texttt{transpose\_matrix} subfunctions in the \texttt{wavefunction.transform} function.

When pivoting is not considered (i.e., $\mathbf{P} = \mathbf{1}$), wave functions are transformed as follows (see Ref.~\cite{Malmqvist1986IJQC} for details).
Suppose that the spin-free MO coefficients are rotated as
\begin{align}
    \tilde{C}_{ri} = \sum_j C_{rj} (\mathbf{L}\mathbf{U})_{ji}
\end{align}
Using $\mathbf{L}$ and $\mathbf{U}$, we compute the matrix elements
\begin{align}
    \mathbf{F} = \mathbf{U}^{-1} - \mathbf{L} - \mathbf{I}
\end{align}
which is performed in the \texttt{process\_matrix} function. It has been shown \cite{Malmqvist1986IJQC} that
this operator can be used to transform the many-body wave functions. Let $\hat{T}$ be the one-particle operator associated with this transformation, i.e.,
\begin{align}
    |\Psi'\rangle = \hat{T}(\mathbf{L}\mathbf{U}) |\Psi\rangle.
\end{align}
We perform
\begin{align}
    |\Psi_{k+1, \sigma}\rangle = (1+ \hat{F}_{k\sigma}) |\Psi_{k,\sigma}\rangle
\end{align}
in which $\hat{F}_{k\sigma}$ is a one-body, spin-orbital operator whose matrix elements are the $k$th column of $\mathbf{F}$, i.e.,
\begin{align}
    \hat{F}_{k\sigma} = \sum_i F_{ik} \hat{a}^\dagger_{i\sigma} \hat{a}_{k\sigma}
\end{align}
This operation is performed recursively first for $\sigma=\alpha$ from $k=0$ ($|\Psi_{0,\alpha}\rangle = |\Psi\rangle$) to $k=n-1$ where $n=\dim(X)$ using a specialized \texttt{apply} function for
"one-column" one-particle operator, followed by the same procedure for $\sigma=\beta$ from  $k=0$ ($|\Psi_{0,\beta}\rangle = |\Psi_{n,\alpha}\rangle$) to $k=n-1$.
The resulting wave function, $|\Psi'\rangle = |\Psi_{n,\beta}\rangle$,
then has the same determinant expansion coefficients $C_{I(\phi)}$ as the coefficients of $|\Psi\rangle$ ($C_{I(\phi')}$) when expressed in the determinants of the new orbital basis.

In the context of exact evolution with a quadratic Hermitian operator, the orbital transformation is performed to
diagonalize the quadratic operator. Assume that the operator is $\hat{A} = \sum_{ij} A_{ij} \hat{E}_{ij}$ with $\hat{E}_{ij} =\sum_\sigma a^\dag_{i\sigma} a_{j\sigma}$;
we diagonalize it so that $\mathbf{X}^\dagger\mathbf{A}\mathbf{X} = \mathbf{a}$ where $\mathbf{a}$ is a diagonal matrix. Note that $\mathbf{X}$ is unitary.
Using the column-pivoted LU decomposition (Eq.~\ref{column}), the propagation can be performed as
\begin{align}
    \exp(-i\hat{A}t) |\Phi\rangle &= \hat{T}((\mathbf{LU})^\dagger) \exp(-i\hat{a}'t) \hat{T}(\mathbf{LU}) |\Phi\rangle
\end{align}
where $\hat{a}'$ is a diagonal operator,
\begin{align}
    \hat{a}' = \sum_{i} (\mathbf{P} \mathbf{a} \mathbf{P}^T)_{ii} \sum_\sigma\hat{a}^\dagger_{i\sigma} \hat{a}_{i\sigma} 
\end{align}
To summarize, in this column-pivoted formulation, the pivot matrix $\mathbf{P}$ is used only to reorder the orbitals in the diagonal operator.
This is more trivial and efficient than reordering the orbitals in the wave functions.

\end{document}